\newcommand{\flux}{\hbox{erg~cm$^{-2}$~s$^{-1}$}}
\newcommand{\lumin}{{erg~s$^{-1}$}}

\newcommand{\msun}{\hbox{${M}_{\odot}$}}

\newcommand{\simgt}{\lower 2pt \hbox{$\, \buildrel {\scriptstyle >}\over {\scriptstyle\sim}\,$}}
\newcommand{\simlt}{\lower 2pt \hbox{$\, \buildrel {\scriptstyle <}\over {\scriptstyle\sim}\,$}}

\newcommand{\chandra}{{\emph{Chandra}}}

\newcommand{\rosat}{{\emph{ROSAT}}}
\newcommand{\einstein}{{\emph{Einstein}}}

\def\aj{AJ}
\def\apj{ApJ}
\def\apjl{ApJL}
\def\apjs{ApJS}
\def\aap{A\&AS}
\def\mnras{MNRAS}
\def\pasp{PASP}
\def\araa{ARA\&A}

\documentclass{an}
\usepackage{graphicx}
\usepackage{times}
\Volume{01}              
\Year{2002}              
\Month{10}               
\Pagespan{223}{227}      

\begin{document}
\lhead[\thepage]{Hornschemeier et~al.: The Weak Outnumbering the Mighty: Normal Galaxies in Deep Chandra Surveys}
\rhead[Astron. Nachr./AN~{\bf XXX} (200X) X]{\thepage}
\headnote{Astron. Nachr./AN {\bf 32X} (200X) X, XXX--XXX}

\title{The Weak Outnumbering the Mighty:\\
Normal Galaxies in Deep Chandra Surveys}

\author{A. E. Hornschemeier,\inst{1}
  F. E. Bauer, \inst{2}
  D. M.~Alexander, \inst{2}
  W. N.~Brandt, \inst{2} \\
  W. L. W.~Sargent, \inst{3}
  C.~Vignali, \inst{2}
  G. P.~Garmire \inst{2}
\and  D. P.~Schneider\inst{2} }
  
\institute{$Chandra$ Fellow, 
Johns Hopkins University, Department of Physics and Astronomy, 3400 N. Charles Street, Baltimore, MD 21218 USA
\and
The Pennsylvania State University, Department of Astronomy and Astrophysics, 525 Davey Lab, University Park, PA 16802 USA
\and
Palomar Observatory, California Institute of Technology,
Pasadena, CA 91125 USA}

\date{Received {\it date will be inserted by the editor};
accepted {\it date will be inserted by the editor}}

\abstract{$Chandra$ is detecting a 
significant population of normal
and starburst galaxies in extremely deep X-ray exposures. For example,  
approximately 15\% of the sources arising in the 2 Ms Chandra Deep Field-North 
survey are fairly normal galaxies, where ``normal" means ``Milky Way-type" 
X-ray emission rather than simply exhibiting an ``optically normal" spectrum.  
Many of these galaxies are being detected at large 
look-back times ($z\approx0.1$--$0.5$),  allowing the study of the evolution of X-ray binary 
populations over significant cosmological timescales.  We are also
detecting individual off-nuclear ultraluminous X-ray sources
(e.g., X-ray binaries), providing the first direct
constraints on the prevalence of lower-mass black holes 
at significantly earlier times. The X-ray emission from
such ``normal" galaxies may also be a useful star-formation rate
indicator, based on radio/X-ray cross-identifications.
We describe the contribution of normal galaxies to the
populations which make up the X-ray background and present their directly
measured X-ray number counts.  We find
that normal and starburst galaxies should dominate the 0.5--2~keV number
counts at X-ray fluxes fainter than 
$\approx 7\times10^{-18}$~erg~cm$^{-2}$~s$^{-1}$ (thus they will outnumber the
``mighty" AGN).
Finally, we look to the future, suggesting that
it is important that the population of X-ray faint normal and
starburst galaxies be well constrained in order to design the
next generation of X-ray observatories.
}

\correspondence{annh@pha.jhu.edu}

\maketitle

\section{Introduction}

The first detection of X-rays from a galaxy outside the Milky Way (the 
Large Magellanic Cloud; Mark et al. 1969) occurred during a sounding rocket 
flight that was similar to the one which discovered the cosmic 
X-ray background (Giacconi et al. 1962).  Approximately ten years later,  
as \einstein\ was being launched, there were still only four normal galaxies detected
in X-rays (the Milky Way, M31, and the two Magellanic Clouds; Helfand et al. 1984).  
Another decade later, the study of normal galaxies in the X-ray band
had become rich enough to warrant an extensive review (Fabbiano 1989), and by
the time of the $Chandra$ and XMM-$Newton$ launches in 1999 it would have been 
difficult to even attempt to list the wealth of X-ray studies of normal
and starburst galaxies as there are so many (but a nice 
example from $ROSAT$ is Read, Ponman, \& Strickland 1997).  

In general, these studies
were confined to the relatively nearby Universe (within $\approx 100$~Mpc).
This paper outlines some of the progress we are now enjoying with the
new capabilities of $Chandra$, which has allowed us to
reach significantly farther.  As will be discussed, there will be
plenty of puzzles to keep us busy well into and {\it past}
the end of the fourth decade since the first X-ray detection of
the LMC.
 
\begin{figure}
\resizebox{\hsize}{!}
{\includegraphics[bbllx=30pt,bblly=003pt,bburx=522pt,bbury=455pt]{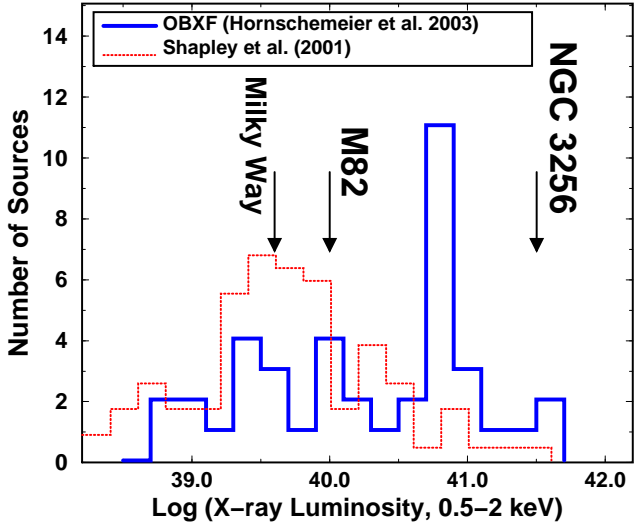}}
\vskip-0.8cm
\caption{
 Histogram of 0.5--2~keV luminosity for OBXF galaxies
in the CDF-N (Hornschemeier et al. 2003).
 The dotted histogram shows the distribution of
late-type galaxies from Shapley, Fabbiano, \&  Eskridge (2001).
The histograms have been offset slightly for clarity.
Also plotted are the 0.5--2~keV X-ray luminosities
of the Milky Way (Warwick 2002), M82 (Griffiths et al. 2000) and NGC~3256
(Moran, Lehnert, \& Helfand 1999).  }
\label{Lxhisto} 
\end{figure}

In 2000, as the Chandra Deep Field  surveys (hereafter CDF-N and CDF-S) 
reached soft X-ray depths 
significantly beyond the $ROSAT$-era deep fields (e.g., Hasinger et al. 1998),
 a population of normal galaxies were found to arise 
(e.g., Hornschemeier et al. 2001; Tozzi et~al. 2001). 
These normal galaxies were found to have X-ray-to-optical flux
ratios lower than that of AGN 
[e.g., $\log {({{f_{\rm X}}\over{f_{\rm R}}})} < -1$].
Recent studies have shown the majority of X-ray sources with 
$-1 \simlt \log{({{f_{\rm X}}\over{f_{\rm R}}})} \simlt -2$ 
to be consistent with infrared and radio-emitting starburst
galaxies (Alexander et al. 2002; Bauer et al. 2002). 
This paper focuses on extragalactic X-ray sources with
even lower X-ray-to-optical flux ratios 
[i.e., $\log{({{f_{\rm X}}\over{f_{\rm R}}})} \simlt -2$], which 
are referred to as ``optically bright, X-ray faint" 
(OBXF) throughout this paper.  The OBXF sources have 
optically normal spectra and 
low X-ray luminosities ($L_{\rm X} \simlt 10^{39}$--$10^{41}$~erg~s$^{-1}$,
0.5--2~keV, see Figure~\ref{Lxhisto}), indicating
they are not obviously dominated by luminous AGN; however,
LLAGNs may be present in some sources.
In general, these sources
are consistent with more quiescent galaxies
(as opposed to starbursts; Hornschemeier et al. 2003) and
comprise an appreciable fraction of the 
faint X-ray source population 
($\approx 30$\% 
of sources having $\approx10^{-17} < f_{\rm X} < 10^{-16}$~erg~cm$^{-2}$~s$^{-1}$, 
0.5--2~keV).   The majority have host galaxies 
of spiral and/or irregular optical morphology and 
$\simgt97$\% show no obvious signs of extended X-ray emission. 

It is important to note that the OBXF sources are distinct 
from the X-ray luminous ``optically normal" galaxies also being discovered
in both deep and moderately-deep X-ray surveys.  These 
galaxies, the prototype of which is the object 
P3 of the HELLAS2XMM survey (e.g., Comastri et al. 2002), 
do not show signatures for AGN in moderate-quality optical
spectra but their X-ray properties suggest
they are powerful AGN.  For further discussion see Moran et al. (2002) and
Comastri et al. in this issue of AN.

The OBXF sources (we use this term somewhat interchangeably with ``normal
galaxy") are being detected at much larger distances 
(500--3000 Mpc; $z\approx0.1$--$0.5$) than was possible for normal galaxies
before $Chandra$, enabling study of 
the cosmological evolution of the X-ray emission 
from galaxies for the first time (e.g., Ptak et al. 2001; Brandt et al. 2001b; 
Hornschemeier et al. 2002; Nandra et al. 2002).  This paper focuses on normal
and starburst galaxies in the CDF-N survey (Brandt et al. 2001a; D.M.~Alexander et al., 
in preparation).  

\begin{figure}
\resizebox{\hsize}{!}
{\includegraphics[bbllx=5pt,bblly=003pt,bburx=522pt,bbury=455pt]{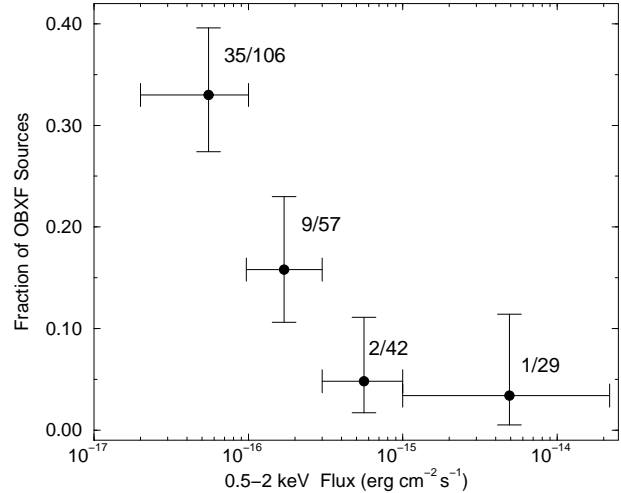}}
\vskip-0.5cm
\caption{
Fraction of X-ray sources that are OBXF as a function of soft-band flux for 
the 2 Ms CDF-N survey (adapted
from Hornschemeier et al. 2003).   This plot is derived from the ``high
exposure area" which covers $\approx40$\% of the CDF-N field.
The number of OBXF sources (numerator) and
total X-ray sources (denominator) are indicated for each bin.
For explanation of the error bars, see Hornschemeier et al.  There is
clearly a large fractional gain in the numbers of fairly normal galaxies with 
increasing X-ray depth.
}
\label{OBXFfraction}
\end{figure}

\section{Source Types and Number Counts}

Even with 1--2~Ms of $Chandra$ data, AGN make up 
the majority of the X-ray sources in deep surveys (the fraction 
is $\approx 60$--75\%, e.g., Rosati et~al. 2002; Barger et~al. 2002; 
Hornschemeier et~al. 2003).  While overall the fraction of 
2~Ms X-ray sources which are OBXF is $\approx 15$\%, the fraction 
increases significantly at faint X-ray fluxes (see \S 1).  
Correspondingly, the number of OBXF sources with X-ray fluxes 
slightly below the 1~Ms detection limit is significant; 
in the CDF-N the number of OBXF sources over the 183 square arcminute
``high exposure area" doubled\footnote{At large off-axis angles, the 2~Ms CDF-N data
are background-limited.  
The gain in number of galaxies quoted is thus pessimistic with respect to
the galaxy number counts. } when the $Chandra$ 
integration time was increased from 1~Ms to 2~Ms (Hornschemeier et~al. 2003).  The gain in the number
of X-ray detected galaxies with increasing X-ray depth is shown in 
Figure~\ref{OBXFfraction}.

\begin{figure}
\resizebox{\hsize}{!}
{\includegraphics[scale=0.90,angle=0]{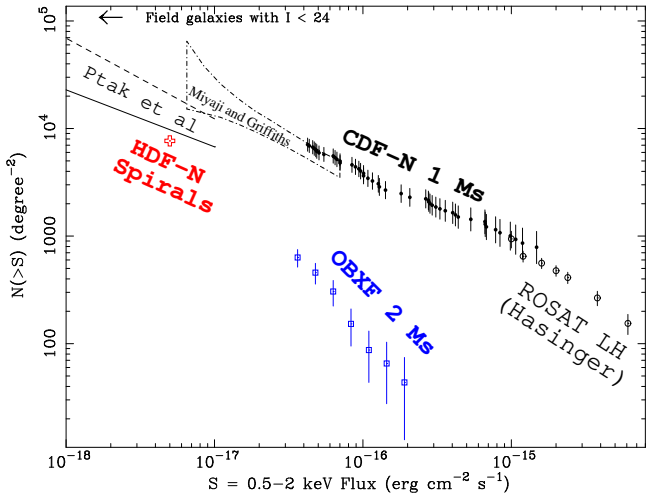}}
\caption{
Number counts for the extragalactic OBXF population.
The CDF-N 1~Ms data are from
Brandt et~al. (2001a). The \rosat\ data are from Hasinger et~al. (1998).
The dashed and solid lines at faint X-ray fluxes show two
predictions of the galaxy number counts made by Ptak et~al. (2001).
The dot-dashed lines mark the results of fluctuation analyses by
Miyaji \& Griffiths (2002).  The cross marks the constraint
from the 1~Ms stacking analysis of Hornschemeier et~al. (2002)
for relatively nearby spiral galaxies ($z\simlt1.5$).
The leftward-pointing arrow indicates the number density of field galaxies at $I=24$.}
\label{LogN}
\end{figure}

We can compare the number counts of the extragalactic
OBXF population with those of the full CDF-N X-ray source population
(see Figure~\ref{LogN} and detailed discussion in Hornschemeier et~al. 2003).
A maximum likelihood fit to the soft-band differential number
counts from $4.2\times10^{-17} $~erg~cm$^{-2}$~s$^{-1}$ to
$2.5\times10^{-16} $~erg~cm$^{-2}$~s$^{-1}$ yields
a slope of $\approx-1.5$  for the corresponding cumulative number counts.  
By comparison, the slope for the general soft-band detected
X-ray source population is quite flat over the same flux range
at $-0.67\pm0.14$  (Brandt et~al. 2001a).

Indirect measures of galaxy number counts, which have been 
able to probe galaxies statistically beyond the formal 1~Ms detection limit, 
have included the stacking analysis work of Hornschemeier et~al. (2002), which
focused on quiescent spiral galaxies, and the fluctuation 
analysis work of Miyaji \& Griffiths (2002). In Figure~\ref{LogN} 
we show these results; an extrapolation of the OBXF 
galaxy counts should intercept the Miyaji fluctuation analysis ``fish"
at a 0.5--2~keV flux of $\approx7\times10^{-18}$~\flux.     This is
a coarse estimate of the flux where the X-ray number counts will
be dominated by normal galaxies and is in reasonable agreement
with the estimates of Ptak et~al. (2001) based on the optical properties
 of field galaxies (also shown in Figure~\ref{LogN}) and with the
stacking analyses of Hornschemeier et~al. (2002).

\vskip-0.5cm

\section{X-ray Emission as a Cosmic SFR Probe}

X-rays provide a unique window into star-formation processes.  Of course,
the penetrating power of hard X-ray emission provides a useful 
cross-check for methods that are  
sensitive to dust obscuration (e.g., ultraviolet emission; 
see the discussion in Seibert et~al. 2002 and references therein), but
studies of X-ray emission also provide important information 
on the evolution of stellar endpoints and hot gas 
that cannot be obtained at any other wavelength (e.g., the multitude
of recent $Chandra$ and XMM-$Newton$ studies of local galaxies).
 Therefore, X-ray emission is not only a possible 
star-formation rate (SFR) indicator 
but an astrophysically compelling way to study galaxies.    

A number of groups have been investigating the use of hard X-rays as 
a surrogate for other star-formation diagnostics in vigorously
star-forming galaxies (e.g., Ranalli et~al. 2002;
Bauer et al. 2002; Nandra et al. 2002).  
It is thought that the X-ray emission of vigorous starbursts
is dominated by the relatively
short-lived high-mass X-ray binary (HMXB) systems, which due to their
relatively short lifetimes ($\sim 10^7$ years) should closely track 
vigorous star formation episodes (e.g., Grimm et al. 2002). 
This use of X-rays as a star-formation diagnostic has been extended to
high redshift in work done on the $z\simgt3$ Lyman Break galaxies 
(e.g., Steidel et~al. 1996).  These galaxies
have been found to exhibit X-ray properties that are broadly similar 
to local vigorous starbursts (e.g., Brandt et~al. 2001b; Nandra et~al. 2002).  
Further work demonstrated that X-ray emission provides a 
test for ultraviolet extinction-correction methods in calculating the  
star-formation rate (Seibert et~al. 2002).  

On longer timescales ($> 10^9$ yr) after a starburst, 
low-mass X-ray binary (LMXBs) evolve to an accreting phase
and emit X-rays. The increased star-formation rate at 
$z\approx1.5$--3 (e.g., Madau et al. 1996) should thus result 
in elevated X-ray emission from LMXB systems in quiescent galaxies at 
$z\approx0.5$--1.  X-ray emission thus represents a ``fossil record" 
of past epochs of star formation 
(e.g., Ghosh \& White 2001; Ptak et~al. 2001), and measurements
of the X-ray luminosities of galaxies can constrain models of 
X-ray binary production.
While X-ray emission from individual quiescent galaxies is not easily detected at 
$z \approx 1$, constraints have been placed on its evolution using stacking 
analyses (e.g., Hornschemeier et~al. 2002).  It was found that at $z\approx1$,
the X-ray luminosity of spiral galaxies was at most a factor of two higher than at
the current epoch.   This has possible implications
for the evolutionary timescale of lower-mass X-ray binaries and will be
improved with larger samples of galaxies.

\section{Lower-mass Black Holes 1 Billion Years Ago}

\vskip-0.09in

Several OBXF sources have X-ray emission offset from the optical galaxy
center.  They have full-band X-ray luminosities $\simgt 10^{39}$~\lumin, 
indicating that they are members of the off-nuclear ultraluminous 
X-ray (ULX) population.  ULX sources have X-ray luminosities in excess 
of that expected for spherically symmetric Eddington-limited accretion 
onto ``stellar" mass (5--20~\msun) black holes. 
  These sources may still be consistent with stellar mass black holes, possibly
representing an unstable, beamed phase in normal high-mass X-ray
binary (HMXB)  evolution (e.g. King et~al. 2001) or harboring  
the most rapidly spinning Kerr black holes among HMXBs 
(e.g., Makishima et~al. 2000). They may also represent a class of
intermediate mass black holes ($\approx 500$--1000~\msun, e.g., Colbert et~al. 2002)
or ultraluminous supernova remnants (e.g., Blair et~al. 2001).   None of the
ULX sources demonstrate spatial extent in the X-ray band but the physical constraints
are not strong due to the low number of counts.

\begin{figure}
\resizebox{\hsize}{!}
{\includegraphics[bbllx=-50pt,bblly=003pt,bburx=382pt,bbury=295pt]{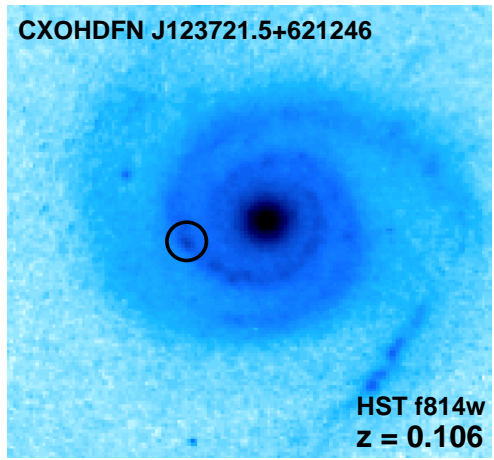}}
\caption{$HST$ image of an off-nuclear X-ray source in the CDF-N 
(from Hornschemeier et al. 2003).
}
\label{ULXfig}
\end{figure}

An example of this type of source is shown in Figure~\ref{ULXfig}. 
We find that $\approx20$\% of galaxies having $M_{B} < -19$ in the CDF-N
field harbor these off-nuclear ULX sources 
(Hornschemeier et~al. 2003).  Variability testing shows that 
some of the CDF-N ULX sources (found mainly near $z\approx0.1$) 
are likely black hole candidates.
 The ULX fraction measured here is only a lower limit; 
even \chandra's sub-arcsecond spatial resolution often 
cannot resolve sources within the central \hbox{$\approx 1$--2~kpc}
 of the nucleus.
Offsets of $\sim1$~kpc are not expected if the object is a supermassive
black hole, but have been found for ULX sources in the local Universe
(see the discussion in Colbert et~al. 2002).  We have thus made the
first pass at evaluating the prevalence of lower-mass black holes at a time 
when the Universe was appreciably younger.  

\section{The Future: Beyond 2 Ms}

Given that normal galaxies are expected to be the majority of
the X-ray sources throughout the Universe, we consider what might be seen 
if we were to look even deeper with $Chandra$.  Extremely long
X-ray observations {\it are} feasible; the 2~Ms CDF-N background level 
indicates $Chandra$ ACIS will
remain approximately photon-limited out to 5~Ms (see also the discussion in this
issue of AN by Alexander et~al.) over the inner $\approx 5^{\prime}$
of the field (or roughly the region of the original 
Hubble Deep Field-North plus Hubble Flanking Fields).
In $\approx 5$~Ms, we should individually detect $\simgt 300$ normal
galaxies in this region and be able to place important statistical 
constraints on thousands of other galaxies.

Large numbers of individual galaxy detections in the X-ray band
will allow construction of galaxy luminosity functions and 
determination of the detailed relationship between the evolution of
X-ray emission and that of the cosmic SFR.  An ultradeep X-ray survey
would also allow an unbiased determination of the frequency of 
ULX sources up to \hbox{$z\approx0.3$--0.4}. 

Beyond X-ray studies of galaxies, {\it any} science which
requires X-ray imaging at fluxes 
$\simlt 7 \times 10^{-18}$~erg~cm$^{-2}$~s$^{-1}$ (0.5--2~keV) may need to
contend with confusion from galaxies if the spatial resolution is not better
than $\approx 2.0^{\prime \prime}$.  Missions such as $XEUS$ are 
planned to reach 
these extremely faint X-ray fluxes, possibly at this fairly 
large spatial resolution.   In order to ensure that we have 
sharp enough vision to operate at these low X-ray fluxes,
it is essential to understand the normal
galaxy population much better.

\nocite{davoXI}
\nocite{BargerCatalog2002,GrimmSFR,Ranalli02,Rosati02}
\nocite{Hornstacking,Tozzi01,Steidel96,Ghosh01,Madau96,Read97,Mark69,Helfand84,Giacconi62,Fabbiano89}
\nocite{Moran02,Comastri02,BrandtLyBreak,Nandra02}
\nocite{Horn01,HornOBXF}
\nocite{Seibert02}
\nocite{King01,Makishima00,Colbert2002,Blair01}
\nocite{Warwick02,Griffiths00,Moran99,Shapley01}
\nocite{BrandtCatalog,Hasinger98,Ptak01,Miyaji02,Hornstacking}
\nocite{Bauerradio,Shapley01}

\acknowledgements

We thank Omar Almaini for the spirited discussion
at the Santander meeting which contributed to this paper.
We gratefully acknowledge the financial support of
NASA grant NAS~8-38252,
$Chandra$ fellowship grant PF2-30021, 
NSF CAREER award AST-9983783,
\chandra\ X-ray Center grant G02-3187A
and NSF grant AST99-00703.


\end{document}